# Electronic properties of new low-temperature superconductors: 3.3K $(Ni_2P_2)(Sr_4Sc_2O_6)$ and 2.7K $(Ni_2As_2)(Sr_4Sc_2O_6)$ from first principles

I. R. Shein*, D. V. Suetin, A. L. Ivanovskii

Institute of Solid State Chemistry, Ural Branch, Russian Academy of Sciences, 91, Pervomaiskaya ul., Ekaterinburg, 620990 Russia

A B S T R A C T

Based on first-principle FLAPW-GGA calculations, we have investigated structural and electronic properties of the recently synthesized tetragonal (space group *P*4/*nmm*) nickel-based pnictide oxide superconductors: 3.3K $(Ni_2P_2)(Sr_4Sc_2O_6)$ and 2.7K $(Ni_2As_2)(Sr_4Sc_2O_6)$. Optimized structural data, electronic bands, total and partial densities of states, and Fermi surface topology have been obtained and discussed in comparison with available experiments and with their Fe-based $(Fe_2P_2)(Sr_4Sc_2O_6)$ and $(Fe_2As_2)(Sr_4Sc_2O_6)$ analogs.

* Corresponding author. Tel.: +7 373 3623115; fax: +7 373 3744495
*E-mail address:* shein@ihim.uran.ru (Shein I.R.)

## 1. Introduction

The discovery of superconductivity in the layered iron pnictides (so-called 1111, 122, and 111 phases) [1-6] and iron chalcogenides Fe–Se(Te) [7] stimulated much activity in search for new related superconductors (SCs), reviews [8-13]. Recently, a new group of more complex five-component phases, namely iron pnictide oxides with general stoichiometry $(Fe_2Pn_2)(A_4M_2O_6)$ (where *Pn* are pnictogens, and *A* and *M* are alkaline-earth and transition metals, respectively) has been discovered, and some physical properties of $(Fe_2Pn_2)(A_4M_2O_6)$ phases have been examined both experimentally and theoretically, review [14].

The most interesting features of $(Fe_2Pn_2)(A_4M_2O_6)$ materials are described below. As distinct from the above 1111, 122, and 111 phases these materials exhibit superconductivity without intentional dopants. So, superconductivity was found in $(Fe_2P_2)(Sr_4Sc_2O_6)$ at 17K [15] and in $(Fe_2As_2)(Sr_4V_2O_6)$ at 37K-46K [16,17]. Thus, this family opened up new guidelines for the development of superconducting materials. These new five-component phases adopt an alternating stacking of $(Fe_2Pn_2)$ and perovskite-like $(A_4M_2O_6)$ blocks, where $(Fe_2Pn_2)/(Fe_2Pn_2)$ separation ($l$ ~ 15.5 Å) is the longest among all known Fe-based SCs. Moreover, the oxide blocks $(A_4M_2O_6)$ in these new compounds have *high chemical flexibility* to a large variety of constituent elements – for example, a set of isostructural phases $(Fe_2As_2)(Sr_4M_2O_6)$, where *M* are Sc [15,18], Cr [18-20], V [16,17], and Mg-Ti [21,22] has been prepared. Another remarkable feature of these perovskite-like oxide blocks is *high structural flexibility*: their thickness can be controlled according to the composition and synthesis conditions. In this way very recently a new series of homologous superconductors with $T_c$ to 42K having thicker oxide layers such as $(Fe_2As_2)(Ca_{n+1}(Sc,Ti)_nO_y)$, where $n$ = 3, 4, and 5, and $y \sim 3n$-1, and isostructural ($Fe_2As_2$)($Ca_{n+1}(Mg,Ti)_nO_y$) with $n$=5 was discovered [23-26].

According to available band structure calculations [27-33], a significant contribution in the window around the Fermi level $E_F$ for ($Fe_2Pn_2$)($A_4M_2O_6$) is made by the bands of the ($Fe_2Pn_2$) blocks, which should play an important role in superconductivity.

Thus, the atomic substitutions inside the ($Fe_2Pn_2$) blocks should exert a profound influence on the physical properties of these phases, and, in particular, on their superconductivity.

In this context, very recently discovered nickel-based pnictide oxide superconductors 3.3K ($Ni_2P_2$)($Sr_4Sc_2O_6$) and 2.7K ($Ni_2As_2$)($Sr_4Sc_2O_6$) [34], which are analogs of the above mentioned iron pnictide oxides, seem highly interesting for a further insight into the nature of these systems. Let us note that several nickel-based analogs of layered 1111 and 122 iron pnictides are known (review [10]), as well as very intriguing [35,36] 2.2K SC $La_3Ni_4P_4O_2$ with an unusual asymmetric distribution of so-called charge reservoir blocks around the conducting ($Ni_2P_2$) blocks), which belong to low-temperature SCs with $T_c$ < 5K. The possible reasons are as follows: as distinct from the iron-based materials, their Ni-based relatives can behave as conventional BCS-like electron-phonon superconductors, or the pairing mechanism may be the same as in the iron-based SCs, but the conditions for the Ni-based systems are far from optimum, as in Fe-based systems [10].

In view of these circumstances, we present here a detailed *ab initio* study of recently discovered nickel-based pnictide oxides ($Ni_2P_2$)($Sr_4Sc_2O_6$) and ($Ni_2As_2$)($Sr_4Sc_2O_6$) and focus our attention on their structural, electronic properties and the Fermi surface topology. As a result, the optimized structural data (including atomic positions), energy bands, total and partial densities of states, Fermi surface topology, as well as the inter-atomic bonding picture have been studied and discussed in comparison with available experiments and with their Fe-based ($Fe_2P_2$)($Sr_4Sc_2O_6$) and ($Fe_2As_2$)($Sr_4Sc_2O_6$) relatives.

## 2. Models and computational aspects

Our calculations were carried out by means of the full-potential method with mixed basis APW+lo (FLAPW) implemented in the WIEN2k suite of programs [37]. The generalized gradient approximation (GGA) to exchange-correlation potential in the PBE form [38] was used. The plane-wave expansion was taken up to $R_{MT} \times K_{MAX}$ equal to 7, and the $k$ sampling with 12×12×3 $k$-points in the Brillouin zone was used. The calculations were performed with full-lattice optimizations including the atomic positions. The self-consistent calculations were considered to be converged when the difference in the total energy of the crystal did not exceed 0.01 mRy, the atomic forces did not exceed 1 mRy/a.u., and the difference in the total electronic charge did not exceed 0.001 $e$ as calculated at consecutive steps. The hybridization effects were analyzed using the densities of states (DOSs), which were obtained by the modified tetrahedron method [39], and some peculiarities of inter-atomic bonding picture were visualized by means of charge density maps.

## 3. Results and discussion

### *3.1. Structural properties.*

The newly synthesized SCs ($Ni_2P_2$)($Sr_4Sc_2O_6$) and ($Ni_2As_2$)($Sr_4Sc_2O_6$) adopt tetragonal (space group *P4/nmm*) crystal structures, which consist of alternate stacking of anti-fluorite ($Ni_2Pn_2$) blocks and perovskite-like ($Sr_4Sc_2O_6$) blocks.

The experimentally obtained lattice parameters are *a* = 4.044 Å and *c* = 15.23 Å for $(Ni_2P_2)(Sr_4Sc_2O_6)$ and *a* = 4.078 Å and *c* = 15.41 Å for $(Ni_2As_2)(Sr_4Sc_2O_6)$ [34].

As no atomic coordinates have been reported for these phases, at the first stage the full structural optimization for both $(Ni_2P_2)(Sr_4Sc_2O_6)$ and $(Ni_2As_2)(Sr_4Sc_2O_6)$ was performed both over the lattice parameters and the atomic positions, Table 1. The calculated lattice constants for $(Ni_2P_2)(Sr_4Sc_2O_6)$ and $(Ni_2As_2)(Sr_4Sc_2O_6)$, which are summarized in the Table 2, are in reasonable agreement with the available [34] experiment: the divergences $(a^{calc} - a^{exp})/a^{exp}$ and $(c^{calc} - c^{exp})/c^{exp}$ are 0.006 and 0.008 for $(Ni_2P_2)(Sr_4Sc_2O_6)$, and 0.006 and 0.004 for $(Ni_2As_2)(Sr_4Sc_2O_6)$, respectively. These divergences should be attributed to the well-known overestimation of the lattice parameters within GGA based methods, as well as to the presence of some amount (~ 5%) of secondary phases such as SrP and $Sc_2O_3$ in the synthesized samples [34].

Our results show that when a smaller P atom ($R^{at}$ = 1.30 Å) is replaced by a larger As atom ($R^{at}$ = 1.48 Å) in anti-fluorite $(Ni_2Pn_2)$ blocks, the lattice parameters and the cell volume grow. On the other hand, in the P → As replacement *anisotropic deformation* of the crystal structure takes place: in comparison with $(Ni_2P_2)(Sr_4Sc_2O_6)$, $a^{calc}$ for $(Ni_2As_2)(Sr_4Sc_2O_6)$ increases only by about 0.032 Å *versus* $c^{calc}$, which increases much more - by about 0.11 Å. This *anisotropic deformation* of the crystal structure is related to strong *anisotropy of inter-atomic bonds*, *i.e.* strong bonds inside $(Ni_2Pn_2)$ blocks *versus* relatively weak ionic coupling between adjacent $(Ni_2Pn_2)/(Sr_4Sc_2O_6)$ blocks, see also below. Note that the same effect of *anisotropic deformation* as a result of atomic substitutions was observed for related layered phases [8-10]. Besides, the $(Ni_2As_2)$ blocks are thicker than the $(Ni_2P_2)$ blocks, whereas the $(Sr_4Sc_2O_6)$ blocks in $(Ni_2As_2)(Sr_4Sc_2O_6)$ are compressed as compared with those in $(Ni_2P_2)(Sr_4Sc_2O_6)$: for example the distances between parallel Sr sheets on the opposite sites of blocks $(Sr_4Sc_2O_6)$ for $(Ni_2As_2)(Sr_4Sc_2O_6)$ are 9.844 Å – *versus* 9.930 Å for $(Ni_2P_2)(Sr_4Sc_2O_6)$.

An even more appreciable effect of *anisotropic deformation* was observed for another type of substitutions, when ions $Ni^{2+}$ (in the discussed nickel-based pnictide oxides with the ideal ionic formula $[Ni_2^{2+}Pn_2^{3-}]^{2-}[Sr_4^{2+}Sc_2^{3+}O_6^{2-}]^{2+}$) are replaced by $Fe^{2+}$. Therefore, we shall compare the present results with the structural data for Fe-based relatives: $(Fe_2P_2)(Sr_4Sc_2O_6)$, where $a^{calc}$ = 4.008 Å and $c^{calc}$ = 15.444 Å, and $(Fe_2As_2)(Sr_4Sc_2O_6)$, where $a^{calc}$ = 4.036 Å and $c^{calc}$ = 15.534 Å, which are obtained within the same FLAPW-GGA approach [27]. In this case, for example, when going from $(Ni_2P_2)(Sr_4Sc_2O_6)$ to $(Fe_2P_2)(Sr_4Sc_2O_6)$, and from $(Ni_2As_2)(Sr_4Sc_2O_6)$ to $(Fe_2As_2)(Sr_4Sc_2O_6)$, the inter-layer distance (parameter *c*) increases (by about 0.074 Å - 0.087 Å), while the parameter *a* decreases by about 0.061 Å - 0.065 Å.

Recently, some empirical correlations between crystallographic parameters and superconducting temperature $T_c$, density of states at the Fermi level ($N(E_F)$), and inter-atomic hybridization effects in iron pnictides have been found [40-43], which seem helpful for elucidation of the prospects for related materials as superconductors.

One trend is that the highest $T_c$ is obtained when *Pn*-Fe-*Pn* bond angles (in $FePn_4$ tetrahedrons, as parameters, which characterize hybridization strength) reach the ideal value of 109.47° for the perfect tetrahedron [40]. In our case, $NiPn_4$ tetrahedrons are disordered with two different *Pn*-Ni-*Pn* angles. These angles are far from the above ideal value, namely, 101.79° and 126.26° for $(Ni_2P_2)(Sr_4Sc_2O_6)$ and 104.32° and 120.36° for $(Ni_2As_2)(Sr_4Sc_2O_6)$ and some other low-temperature Ni-based SCs [10].

Another correlation is between the transition temperature $T_c$ (and $N(E_F)$) and the anion height ($\Delta z_a$) with respect to Fe sheet inside ($Fe_2Pn_2$) blocks [41-43]. The theoretical background is that the parameter $\Delta z_a$ may be a switch between high-$T_c$ nodeless and low-$T_c$ nodal pairings for the Fe-based SCs [43]. For these systems, the dependence ($T_c$ *versus* $\Delta z_a$) has a clear maximum at about $\Delta z_a$ = 1.38 Å [41]. Note that the parameter $\Delta z_a$ for 17K ($Fe_2P_2$)($Sr_4Sc_2O_6$) also corresponds to this dependence [41]. For the examined Ni-based phases this parameter is much less: $\Delta z_a$ = 1.03 Å for ($Ni_2P_2$)($Sr_4Sc_2O_6$) and $\Delta z_a$ = 1.18 Å for ($Ni_2As_2$)($Sr_4Sc_2O_6$).

### *3.2. Electronic band structure and Fermi surface*

Figure 1 shows the band structures of ($Ni_2P_2$)($Sr_4Sc_2O_6$) and ($Ni_2As_2$)($Sr_4Sc_2O_6$) as calculated along the high-symmetry *k* lines; the corresponding bandwidths are listed in Table 3.

In the band structure of ($Ni_2P_2$)($Sr_4Sc_2O_6$), two lowest bands lying around -11.5 eV below the Fermi level arise mainly from P 3*s* states and are separated from the near-Fermi valence bands by a gap. The next groups of occupied bands are placed in the intervals from -6.0 eV to -1.0 eV and from - 1.0 eV to $E_F$. The Fermi level is intersected by rather low-dispersive bands responsible for the metallic-like character of this system. The main distinctive features are the quasi-flat bands along the *A-M* and *R-X* high-symmetry directions, which originate mainly from Ni 3*d* electronic states, see also below. These bands form four sheets of the Fermi surface (FS), and are cylindrical-like and parallel to the $k_z$ direction, see Fig. 2. Three of them (electronic-like) are centered along the *A−M* line, whereas the fourth sheet is centered along the *R-X* line and is hole-like.

Thus, our calculations demonstrate an entirely different FS topology for 3.3K $(Ni_2P_2)(Sr_4Sc_2O_6)$ as compared with the isostructural 17K superconductor $(Fe_2P_2)(Sr_4Sc_2O_6)$, for which the Fermi surface is made up of five sheets; three of them are hole-like and are centered along the $\Gamma-Z$ direction, while the other two (electronic-like) sheets are centered along the $A-M$ direction [27]. The Fermi surfaces of the other known Ni-based low-temperature SCs such as $SrNi_2As_2$ ($T_c \sim$ 0.6K), $BaNi_2As_2$ ($T_c \sim$ 0.7K), and $BaNi_2P_2$ ($T_c \sim$ 3 K) are of a multi-sheet three-dimensional type. [44] These differences in FS topology are due mainly to changes in inter-atomic distances and bonding angles in $[Ni_2Pn_2]$ blocks, as well as to the degree of band filling. Note that among the examined Ni-based SCs, LaNiPO [45,46] has the FS topology, which is most similar to that obtained for $(Ni_2P_2)(Sr_4Sc_2O_6)$.

The band structure picture of $(Ni_2As_2)(Sr_4Sc_2O_6)$ is very similar to that for $(Ni_2P_2)(Sr_4Sc_2O_6)$, see Fig. 1 and Table 3, whereas the FS of $(Ni_2As_2)(Sr_4Sc_2O_6)$, Fig. 2, differs sharply from the Fermi surface of $(Fe_2As_2)(Sr_4Sc_2O_6)$ [27], which may be expected from the different electron count.

*3.3. Density of states and inter-atomic bonding.*

The calculated total and partial densities of states (DOSs) of $(Ni_2P_2)(Sr_4Sc_2O_6)$ and $(Ni_2As_2)(Sr_4Sc_2O_6)$ are very similar, see Fig. 3. So, for $(Ni_2P_2)(Sr_4Sc_2O_6)$, the quasi-core peak A is almost completely made up of P 3*s* orbitals with a very small admixture of Sr *s*,*p* orbitals. The next peak B is formed mainly by contributions of Ni 3*d* and P 3*p* states (from $(Ni_2P_2)$ blocks) and O 2*p* states with admixtures of Sc 3*d* and valence Sr s,*p* states from $(Sr_4Sc_2O_6)$ blocks. These results indicate that the Fe–P and Sc–O covalent bonds in the corresponding blocks are due to the hybridization of Fe 3*d* - P 3*p* and Sc 3*d* - O 2*p* states, and the bonding picture for $(Ni_2P_2)(Sr_4Sc_2O_6)$ differs considerably from the ionic model.

The occupied bands near the Fermi level (peak C, Fig. 3) and the lowest conduction bands are formed exclusively by the states of $(Ni_2P_2)$ blocks, with predominant contributions of Ni 3*d* states, whereas the occupied and unoccupied states of atoms forming $(Sr_4Sc_2O_6)$ blocks are separated by a gap. Thus, $(Ni_2P_2)(Sr_4Sc_2O_6)$ should be considered as a quasi-two-dimensional material formed from conducting $(Ni_2P_2)$ blocks, which alternate with insulating $(Sr_4Sc_2O_6)$ blocks. Therefore, conduction in $(Ni_2P_2)(Sr_4Sc_2O_6)$ (and in $(Ni_2As_2)(Sr_4Sc_2O_6)$, see Fig. 3) will be strongly anisotropic, *i.e.* it will happen in $(Ni_2Pn_2)$ blocks, as well as in other related Ni-based low-$T_c$ SCs such as LaNiPO [45,46], $La_3Ni_4P_4O_2$ [36], $ANi_2Pn_2$ [44] or LaNiBiO [47].

Let us also note that the bands at $E_F$ in $(Ni_2Pn_2)(Sr_4Sc_2O_6)$ are more dispersive than in the Fe-based relatives [27]. This leads to higher in-plane Fermi velocity and to lower values of density of states at the Fermi level, $N(E_F)$, see Table 4. For a further insight into the near-Fermi states, in Table 4 we give also the orbital-decomposed density $N(E_F)$. In a tetrahedral crystal field, the Ni 3*d* bands will split into lower lying *e* ($d_{z^2}$, $d_{x^2-y^2}$) and upper lying $t_2$ ($d_{xy}$, $d_{yz}$, $d_{xz}$) orbitals. However, owing to the distortions of $NiPn_4$ tetrahedrons (see above), no separation of the *d* orbitals in energy was found, and all the five Ni 3*d* orbitals as well as three *Pn p* orbitals have comparable contributions to $N(E_F)$, and all these orbitals should contribute to charge carriers at the Fermi level – in contrast to Fe-based SCs, where the near-Fermi bands arise mostly from the $d_{xy}$, $d_{yz}$, $d_{xz}$ states with some admixture of $d_{z^2}$ orbitals [8-11].

The obtained data allow us to estimate the Sommerfeld constants (γ) under assumption of the free electron model as $\gamma = (\pi^2/3)N(E_F)k^2_B$. It is seen from Table 4 that the values of $\gamma$ increase when phosphorus is replaced by arsenic - in contrast to their Fe-based relatives, where the coefficients $\gamma$ decrease as going from $(Fe_2P_2)(Sr_4Sc_2O_6)$ to $(Fe_2As_2)(Sr_4Sc_2O_6)$ [27]. On the other hand, the values of $\gamma$ for $(Ni_2P_2)(Sr_4Sc_2O_6)$ and $(Ni_2As_2)(Sr_4Sc_2O_6)$ are comparable with those for other known Ni-based SCs, for which Sommerfeld coefficients range from 4.35 $mJ/K^2$·mole (for $SrNi_2As_2$) to 9.23 $mJ/K^2$·mole (for $La_3Ni_4Si_4$) [10].

Finally, let us note that one of the most intriguing peculiarities of the Fe-based SCs is the presence of typically metallic collective excitations, such as itinerant magnetization waves in the non-superconducting undoped parent 1111 or 122 phases such as $Ln$FeAsO or $A$$Fe_2As_2$ (where $Ln$ are early rare earth metals La, Ce, Sm, Dy, Gd, and $A$ are alkali earth metals Ca, Sr, and Ba), for which at least three different competing types of magnetic fluctuations have been predicted, reviews [8-13].

On the contrary, the available data [10] give no evidence for magnetism in Ni-based relatives excluding probably $La_3Ni_4P_4O_2$. In our calculations for $(Ni_2P_2)(Sr_4Sc_2O_6)$ and $(Ni_2As_2)(Sr_4Sc_2O_6)$ we find a non-magnetic ground state. In the simple way within our band structure calculations, the magnetic instability may be explained using the Stoner criterion, according to which magnetism may occur if $N(E_F)^* I > 1$. Here, $N(E_F)^*$ is the density of states at the Fermi level on an atom per spin basis. Using the typical value $I = 0.9$ [48], our estimations show that this indicator changes from 0.704 $(Ni_2P_2)(Sr_4Sc_2O_6)$ to 0.866 for $(Ni_2As_2)(Sr_4Sc_2O_6)$, testifying to the non-magnetic behavior of these species.

If we assume, that the examined $(Ni_2As_2)(Sr_4Sc_2O_6)$ and $(Ni_2P_2)(Sr_4Sc_2O_6)$ phases will behave as conventional BCS-like electron-phonon superconductors (like related Ni-based 1111 or 122 phases [10,45]), the simplified correlation $T_c \sim N(E_F)$ may be used for qualitative explanations of variation of $T_c$ for these materials. In our calculations the opposite tendency is found: $N(E_F)\{3.3K$ $(Ni_2P_2)(Sr_4Sc_2O_6)\}$ = 3.13 states/eV/form.unit < $N(E_F)\{2.7K$ $(Ni_2As_2)(Sr_4Sc_2O_6)\}$ = 3.85 states/eV/form.unit, which is due chiefly to the growth of contributions from Ni 3*d* orbitals for $(Ni_2As_2)(Sr_4Sc_2O_6)$, Table 4. Within the Allen-Dynes model [49] ($T_c \sim <\omega>\exp\{f(\lambda_{ep})\}$), where <ω> is the averaged phonon frequency, and $\lambda_{ep}$ is the electron-phonon coupling constant, one of the reasons why $T_c$ of $(Ni_2As_2)(Sr_4Sc_2O_6)$ decreases may be the lowering of <ω> for this As-containing material in comparison with $(Ni_2P_2)(Sr_4Sc_2O_6)$. So, direct calculations [45] of phonon dispersions for related layered LaNiPO demonstrate that the pnictogen atoms actively participate in the formation of all phonon bands. Certainly, for $(Ni_2As_2)(Sr_4Sc_2O_6)$ and $(Ni_2P_2)(Sr_4Sc_2O_6)$ the similar calculations of phonon dispersions and numerical estimations of electron-phonon coupling are necessary for the further explanation of the pairing mechanism for these materials.

Finally, let us discuss the inter-atomic bonding for the considered Ni-based phases. To describe the ionic bonding for these materials, we carried out Bader [50] analysis, and then the effective charges defined as $\Delta Q = Q^B - Q^i$ (where $Q^B$ are so-called Bader charges, and $Q^i$ are formal ionic charges of atoms as obtained from the purely ionic model, which considers the usual oxidation numbers of atoms) are estimated and presented in Table 5. These results show that inside each block: $(Sr_4Sc_2O_6)$ and $(Ni_2(P,As)_2)$, the ionic bonding takes place between the ions with opposite charges: $(Sr^{a+},Sc^{b+})$ – $O^{c-}$ and $Ni^{n+}$ - $(P,As)^{m-}$, but the effective atomic charges ΔQ are much smaller than predicted in the idealized ionic model. Besides, taking into account the distributions of atoms in the above mentioned blocks, it is possible to estimate the charge transfer between these blocks. We found that for $(Ni_2P_2)(Sr_4Sc_2O_6)$ the charge transfer (of about 0.62*e*) occurs from positively charged blocks $(Sr_4Sc_2O_6)$ to negatively charged conducting blocks $(Ni_2P_2)$ and is similar to that in $(Ni_2As_2)(Sr_4Sc_2O_6)$ (about 0.60*e*), whereas between these blocks the ionic bonding takes place.

The charge density distribution (Fig. 4) reveals that (i) the mentioned covalent bonding Ni-(P,As) and Sc-O inside $(Ni_2(P,As)_2)$ and $(Sr_4Sc_2O_6)$ blocks takes place, (ii) some overlapping of Sr orbitals with neighboring oxygen atoms occurs, which is indicative of partially covalent bonding of Sr ions, and (iii) between the adjacent $(Ni_2(P,As)_2)$ and $(Sr_4Sc_2O_6)$ blocks mainly ionic bonds arise owing to $(Ni_2(P,As)_2)$ → $(Sr_4Sc_2O_6)$ charge transfer. Further, inside $(Ni_2(P,As)_2)$ blocks the metallic-like Ni-Ni bonding occurs as a result of overlapping of the near-Fermi Ni 3*d* states.

Thus, summarizing the above results, the inter-atomic bonding for $(Ni_2P_2)(Sr_4Sc_2O_6)$ and $(Ni_2As_2)(Sr_4Sc_2O_6)$ systems can be characterized as a high-anisotropic mixture of ionic, covalent, and metallic contributions, where inside $(Ni_2(P,As)_2)$ blocks mixed covalent-ionic bonds Ni-P(As) take place together with the metallic-like Ni-Ni bonds (owing to delocalized near-Fermi Ni 3*d* states); mixed covalent-ionic inter-atomic bonds arise also inside $(Sr_4Sc_2O_6)$ blocks, whereas inter-blocks bonding is basically of ionic type. A similar picture was found for other layered Ni-based SCs [44,46,47,51].

**4. Conclusions**

In summary, we have presented a detailed *ab initio* study of the structural and electronic properties of the newly synthesized Ni-based low-$T_c$ SCs: $(Ni_2P_2)(Sr_4Sc_2O_6)$ and $(Ni_2As_2)(Sr_4Sc_2O_6)$. These materials can be characterized as quasi-two-dimensional non-magnetic ionic metals formed from conducting $(Ni_2Pn_2)$ blocks, which alternate with insulating $(Sr_4Sc_2O_6)$ blocks; the binding between them is mostly ionic. Their Fermi surfaces are formed by four cylindrical-like sheets parallel to the $k_z$ direction. Three of them (electronic-like) are centered along the $A-M$ line, whereas the fourth sheet is centered along the $R$-$X$ line and is hole-like. The topology of FSs of the examined Ni-based five-component phases differs considerably from isostructural Fe-based relatives $(Fe_2Pn_2)(Sr_4Sc_2O_6)$, as well as from $ANi_2Pn_2$ phases, but is very similar to the Fermi surfaces for other Ni-based layered low-$T_c$ superconductors LaNiPO and LaNiBiO. We also find that the orbital components on the Fermi level are of a mixed type, where all five Ni 3$d$ orbitals as well as three *Pn p* orbitals are presented - in contrast to Fe-based SCs, where the near-Fermi bands are due mostly to the $d_{xy}$, $d_{yz}$, $d_{xz}$ states with some admixture of $d_{z}2$ orbitals.

Clearly, further theoretical and experimental studies are needed to clarify the pairing mechanism in these new materials and to search for related systems. Thus, direct calculations of phonon dispersions and numerical estimations of electron-phonon coupling, as well as studies of electronic correlations [52] will be very helpful. Finally, the known high chemical and structural flexibility of perovskite-like oxide blocks in these five-component phases may be the key for further theoretical and experimental search of new Ni-based materials by replacements of Sc atoms by other $d$ atoms inside the insulating $(Sr_4Sc_2O_6)$ blocks, which may have exciting superconducting and magnetic properties.

**Acknowledgements**

Financial support from the RFBR (Grants 09-03-00946 and 10-03-96008-Ural) is gratefully acknowledged.

**Table 1**
The optimized atomic positions for $(Ni_2P_2)(Sr_4Sc_2O_6)$ and $(Ni_2As_2)(Sr_4Sc_2O_6)$.

| atomic positions | *x* | *y* | *z* * |
|---|---|---|---|
| $Sr_1$ (2c) | ¾ | ¾ | 0.1772 (0.1811) |
| $Sr_2$ (2c) | ¾ | ¾ | 0.4114 ( 0.4120) |
| Sc (2c) | ¼ | ¼ | 0.3019 (0.3039) |
| $O_1$ (4f) | ¼ | ¼ | 0.2797 (0.2824) |
| $O_2$ (2c) | ¼ | ¼ | 0.4311 (0.4324) |
| Ni (2b) | ¼ | ¾ | 0 (0) |
| As (2c) | ¼ | ¼ | 0.0671 (0.0770) |

* the data for $(Ni_2P_2)(Sr_4Sc_2O_6)$ and $((Ni_2As_2)(Sr_4Sc_2O_6))$ are given.

**Table 2**
The optimized lattice parameters (*a* and *c*, in Å) and cell volume (*V*, in $Å^3$) for $(Ni_2P_2)(Sr_4Sc_2O_6)$ and $(Ni_2As_2)(Sr_4Sc_2O_6)$.

| system | *a* | *c* | *V* |
|---|---|---|---|
| $(Ni_2P_2)(Sr_4Sc_2O_6)$ | 4.0687 (4.044) | 15.3563 (15.23) | 254.21 (249.09) |
| $(Ni_2As_2)(Sr_4Sc_2O_6)$ | 4.1008 (4.078) | 15.4636 (15.41) | 260.04 (256.19) |

* available experimental data [34] are given in parentheses.

**Table 3**
Calculated bandwidths (in eV) of occupied states for $(Ni_2P_2)(Sr_4Sc_2O_6)$ and $(Ni_2As_2)(Sr_4Sc_2O_6)$.

| system | Quasi-core Pn s band | Band gap | Valence band (up to $E_F$) |
|---|---|---|---|
| $(Ni_2P_2)(Sr_4Sc_2O_6)$ | 2.03 | 4.08 | 6.11 |
| $(Ni_2As_2)(Sr_4Sc_2O_6)$ | 1.67 | 4.99 | 5.75 |

**Table 4**
Total and partial densities of states at the Fermi level $N(E_F)$, $N^l(E_F)$ (in states/eV/form.unit), and Sommerfeld coefficients $\gamma$ (in mJ/K$^2$·mole) for $(Ni_2P_2)(Sr_4Sc_2O_6)$ and $(Ni_2As_2)(Sr_4Sc_2O_6)$.

| system* | 1 | 2 | system * | 1 | 2 |
|---|---|---|---|---|---|
| $N_{tot}$ | 3.13 | 3.85 | $N^{Ni\,3d}{}_{xy}(E_F)$ | 0.32 | 0.42 |
| $N^{Ni\,3d}(E_F)$ | 1.41 | 1.81 | $N^{Ni\,3d}{}_{xz}{}^{+3d}{}_{yz}(E_F)$ | 0.48 | 0.60 |
| $N^{(As,\,P)p}(E_F)$ | 0.43 | 0.42 | $N^{(As,\,P)p}{}_x$ | 0.14 | 0.15 |
| $N^{Ni\,3d}{}_{z^2}(E_F)$ | 0.22 | 0.33 | $N^{(As,\,P)p}{}_y{}^{+p}{}_z$ | 0.29 | 0.27 |
| $N^{Ni\,3d}{}_{x^2-y^2}(E_F)$ | 0.39 | 0.46 | $\gamma$ | 7.42 | 9.12 |

* 1- $(Ni_2P_2)(Sr_4Sc_2O_6)$; 2- $(Ni_2As_2)(Sr_4Sc_2O_6)$.

**Table 5**
Effective atomic charges ($\Delta Q$, in *e*) for $(Ni_2P_2)(Sr_4Sc_2O_6)$ and $(Ni_2As_2)(Sr_4Sc_2O_6)$.

| system | Ni | P(As) | $Sr_1$ * | $Sr_2$ | Sc | $O_1$ * | $O_2$ |
|---|---|---|---|---|---|---|---|
| $(Ni_2P_2)(Sr_4Sc_2O_6)$ | +1.833 | -2.129 | +0.548 | +0.481 | +1.221 | -0.641 | -0.672 |
| $(Ni_2As_2)(Sr_4Sc_2O_6)$ | +1.976 | -2.284 | +0.546 | +0.500 | +1.234 | -0.644 | -0.679 |

* $Sr_{1,2}$ and $O_{1,2}$ – non-equivalent atoms, see Table 1.

## FIGURES

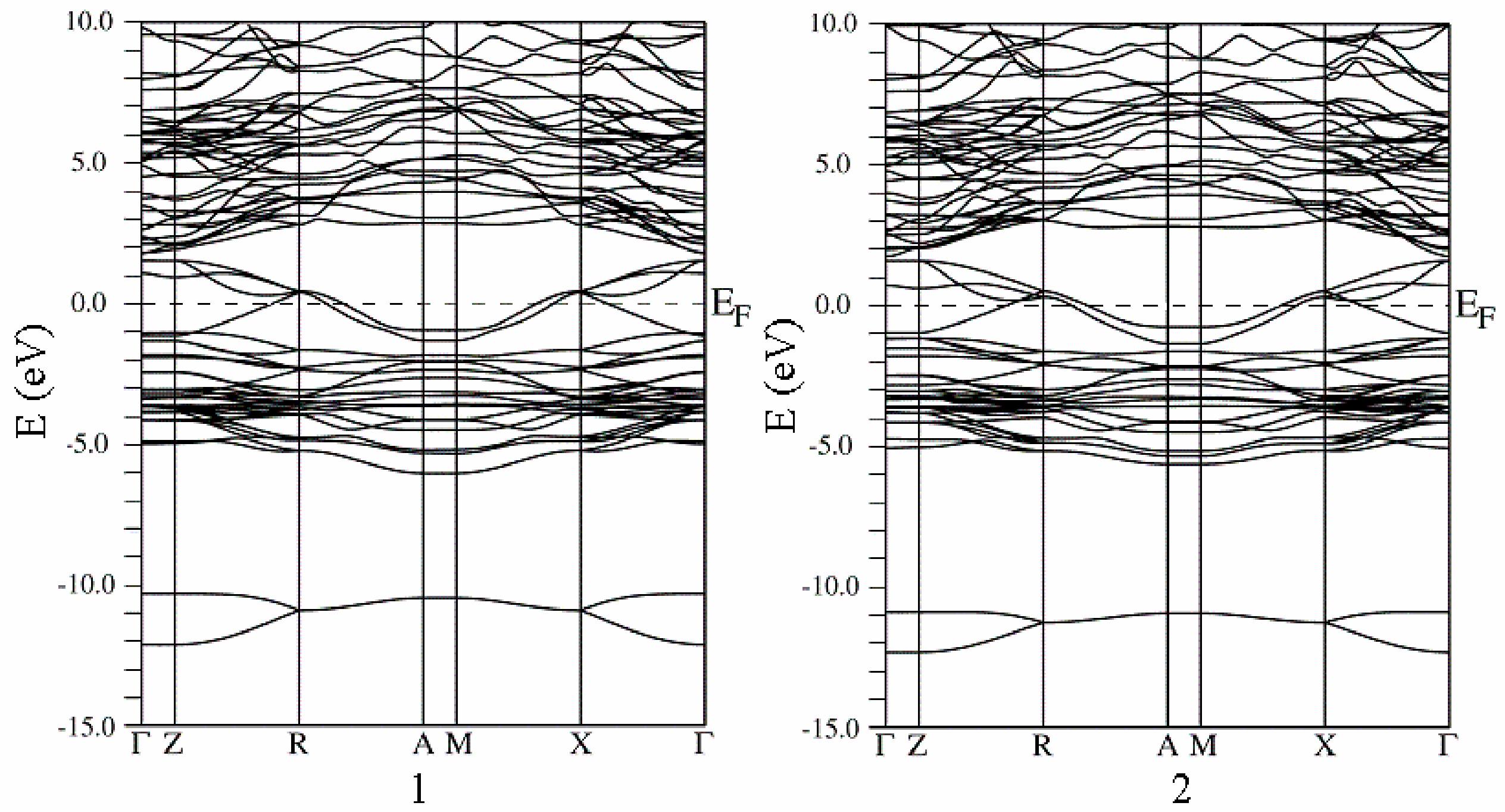

**Fig. 1.** Electronic bands for $(Ni_2P_2)(Sr_4Sc_2O_6)$ (1) and $(Ni_2As_2)(Sr_4Sc_2O_6)$ (2).

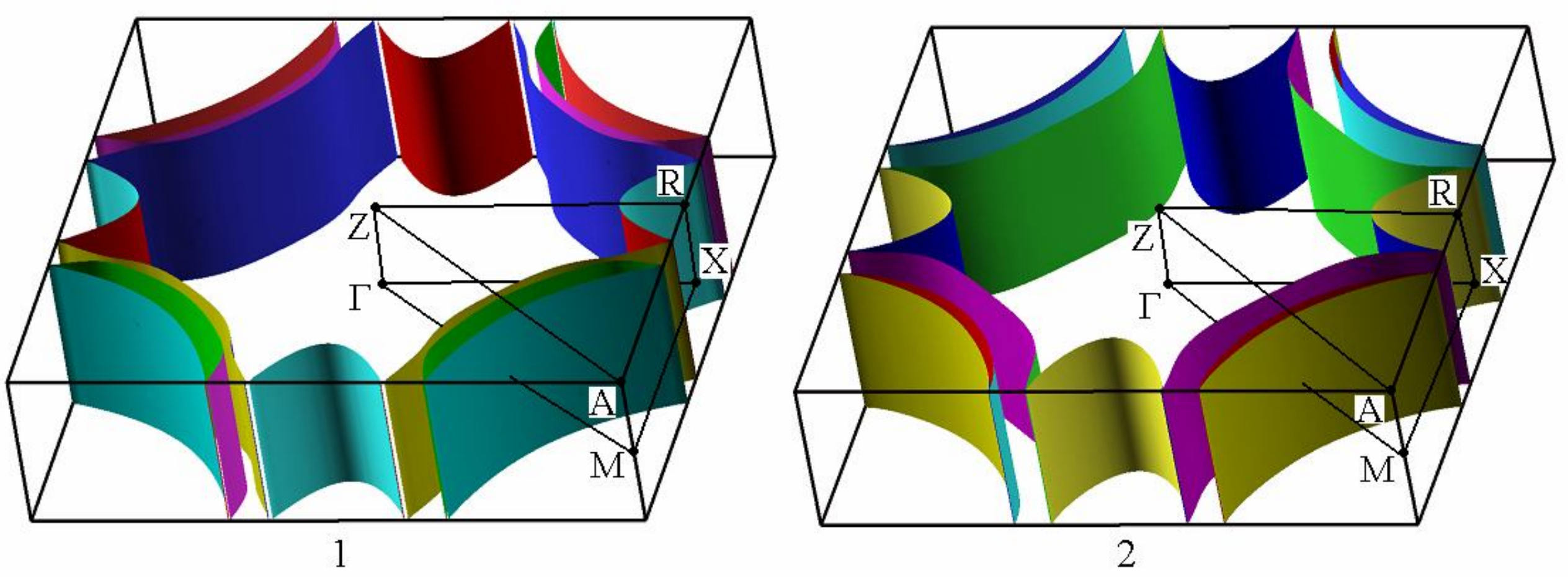

**Fig. 2**. (*Color online*) Fermi surfaces for $(Ni_2P_2)(Sr_4Sc_2O_6)$ (1) and $(Ni_2As_2)(Sr_4Sc_2O_6)$ (2).

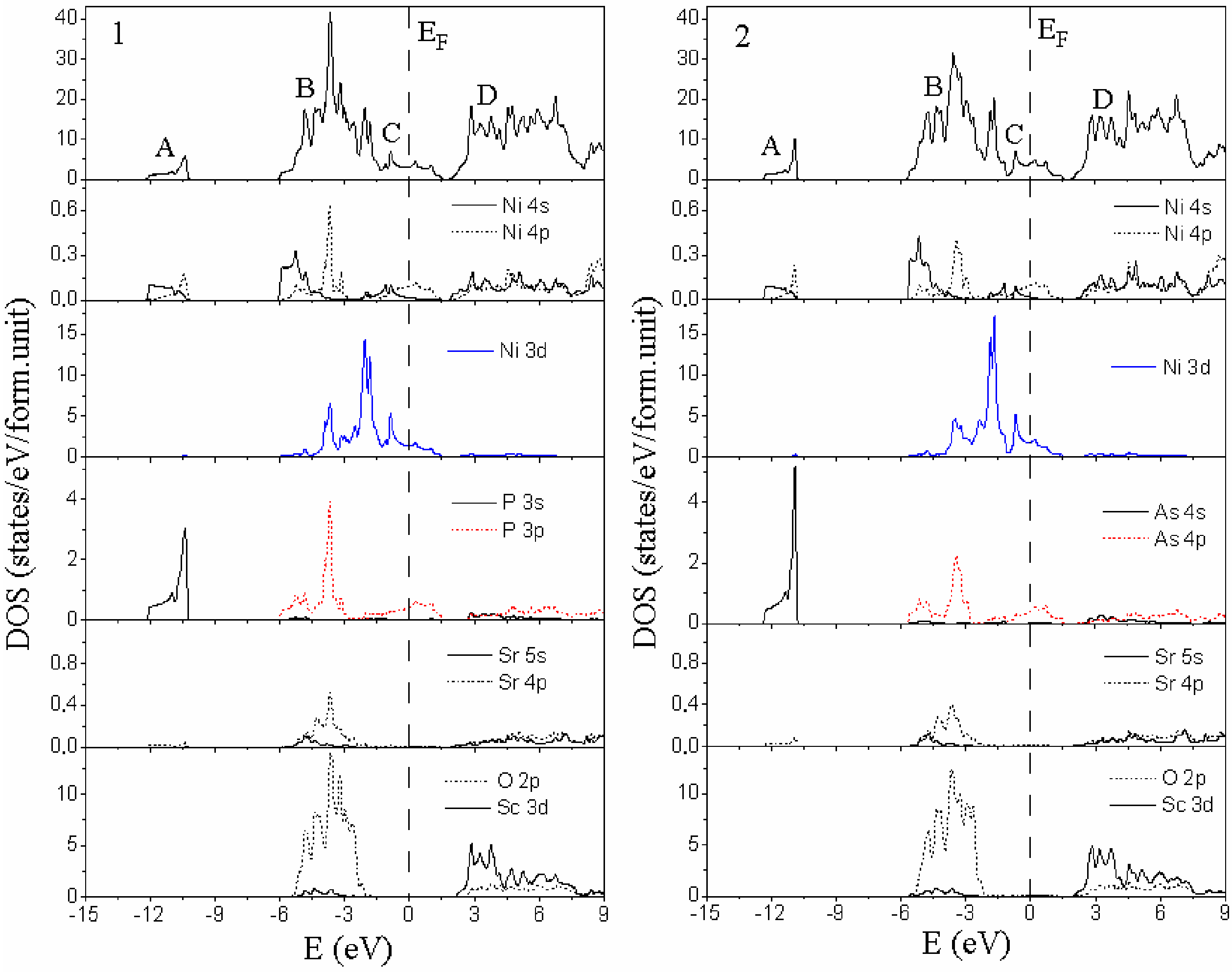

**Fig. 3**. **(***Color online***)** Total (*upper panels*) and partial densities of states (*bottom panels*) for $(Ni_2P_2)(Sr_4Sc_2O_6)$ (1) and $(Ni_2As_2)(Sr_4Sc_2O_6)$ (2).

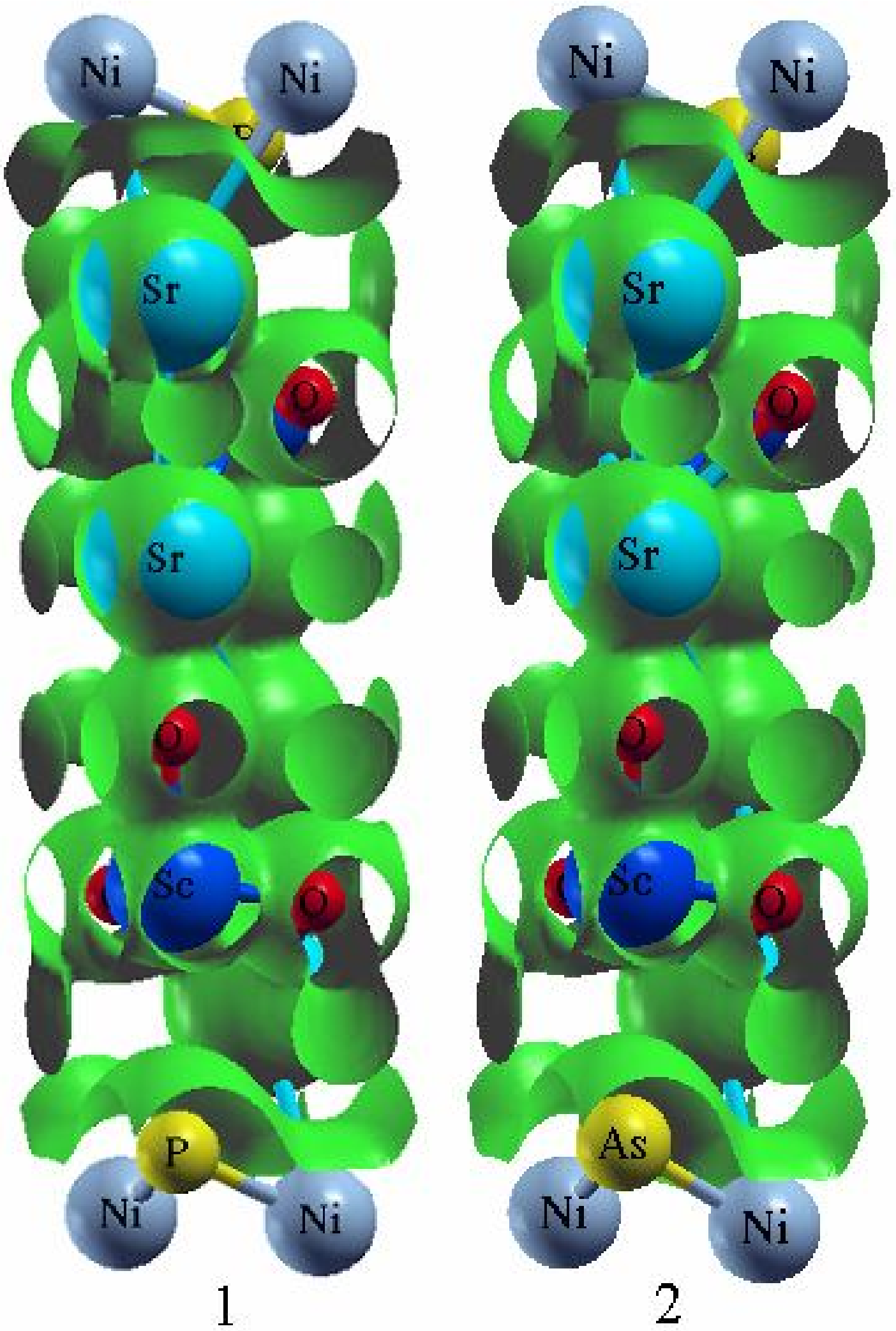

**Fig. 4.** (*Color online*). Charge density iso-surfaces (0.15 e/Å$^3$) of valence states for $(Ni_2P_2)(Sr_4Sc_2O_6)$ (1) and $(Ni_2As_2)(Sr_4Sc_2O_6)$ (2).